\documentclass[runningheads]{llncs}
\usepackage[T1]{fontenc}
\usepackage{graphicx}
\usepackage{subcaption}
\usepackage{algorithm} %コード用
\usepackage{algpseudocode} %コード用
\usepackage{amsmath,amssymb} %4章の見出し
\DeclareMathOperator*{\argmax}{argmax} %３章の数式
\spnewtheorem{prob}{Problem}{\bfseries}{\rmfamily} %3章の見出し
\spnewtheorem{mthd}{Method}{\bfseries}{\rmfamily} %4章の見出し

\begin{document}
\title{Influence Ranking Improvement\\via Link Addition in Social Networks}
\titlerunning{Influence Ranking Improvement via Link Addition}
% If the paper title is too long for the running head, you can set
% an abbreviated paper title here
%
\author{Taiki Ikemoto\inst{1}\and
Sho Tsugawa\inst{2}
}
\authorrunning{T. Ikemoto and S. Tsugawa}
% First names are abbreviated in the running head.
% If there are more than two authors, 'et al.' is used.
%
\institute{Graduate School of Science and Technology, University of Tsukuba\\
1–1–1 Tennodai, Tsukuba, Ibaraki, 305–8573 Japan\\
\email{t.ikemoto@snlab.cs.tsukuba.ac.jp}\and
Institute of Systems and Information Engineering, University of Tsukuba\\
1–1–1 Tennodai, Tsukuba, Ibaraki, 305–8573 Japan\\
\email{s-tugawa@cs.tsukuba.ac.jp}\\
}
\maketitle              % typeset the header of the contribution
\begin{abstract}
Social media platforms increasingly rely on influential users for information dissemination in domains such as marketing and political campaigns.
As the value of being recognized as influential grows, users may have incentives to strategically enhance their influence.
In this study, we investigate whether and to what extent the influence ranking of a target node can be improved by adding a limited number of outgoing links in the context of influence maximization (IM).
We formulate the Link Selection Problem for Influence Ranking Improvement (LSP-IRI) as the problem of selecting a fixed-size set of additional outgoing links from a target node in order to maximize its rank improvement.
To examine this problem, we consider two representative heuristic strategies: a greedy method that directly optimizes rank improvement and a computationally efficient random search method.
We conduct experiments on four real-world networks ranging from thousands to hundreds of thousands of nodes.
The results show that even a few added links can substantially improve IM-based influence rankings.
In particular, the greedy method improves the ranks of nodes initially ranked around 50 by several tens of positions on average with only three added links, sometimes moving them into the top 10.
The random search method achieves smaller gains under strict budgets but reduces computation time by up to approximately 98\% compared with the greedy method and becomes effective when larger budgets are allowed.
These findings show that IM-based influence rankings are sensitive to limited local structural modifications and highlight a trade-off between ranking improvement and computational efficiency.


\end{abstract}
\section{Introduction}

The global rise of influencer marketing, which leverages influential users to promote products and services, has made algorithmic assessments of user influence increasingly consequential.
Because influential individuals play a critical role in shaping consumer awareness and purchasing behavior \cite{barari2025influence}, social media platforms have evolved into competitive environments where user visibility and influence translate directly into economic and social capital.
As recognition as an {\em influencer} becomes increasingly associated with professional and social success, users may have incentives to strategically enhance their perceived influence \cite{fetter2023desires}.
This trend highlights the importance of influence-ranking algorithms, not only as tools for identifying influential users, but also as algorithmic mechanisms whose robustness must be carefully examined. If small changes in the network structure can substantially alter influence rankings, then influence-based selection mechanisms may be more vulnerable to strategic behavior than previously understood.

While extensive research has focused on identifying influencers from a platform-centric perspective, relatively little attention has been paid to the perspective of individual users who seek to be selected as influencers.
Among existing approaches, influence maximization (IM) is one of the most widely studied frameworks for identifying influential seed nodes \cite{kempe2003maximizing}.
Given a network and a diffusion model, IM algorithms aim to identify a set of seed nodes that maximizes the expected influence spread during information propagation.
A substantial body of work has developed efficient and scalable algorithms capable of identifying such highly influential nodes, even in large-scale networks. 
Recent studies have further extended IM to practical settings, including fairness-aware influencer recommendation and multi-objective influence maximization \cite{dam2025fair2vec,biswas2022moimp}.
However, these studies primarily address the platform’s objective of maximizing global reach and typically assume a fixed network structure. In contrast, real social networks are dynamic, and users can modify their local connectivity through link formation.
Despite its importance, little is known about how susceptible IM-based influence rankings are to such small structural changes.
This gap raises a fundamental research question: how vulnerable are influence rankings to limited link additions, and to what extent can the ranking of a target node be improved through local network interventions?
Conceptually, this problem differs from conventional IM because the IM algorithm is treated as a ranking mechanism rather than as an optimizer of a platform-wide seed set. It also differs from link recommendation because the objective is not to predict likely links but to select links according to their effect on the target node’s IM-based rank.

\begin{figure}[t]
\centering
\includegraphics[width=\linewidth]{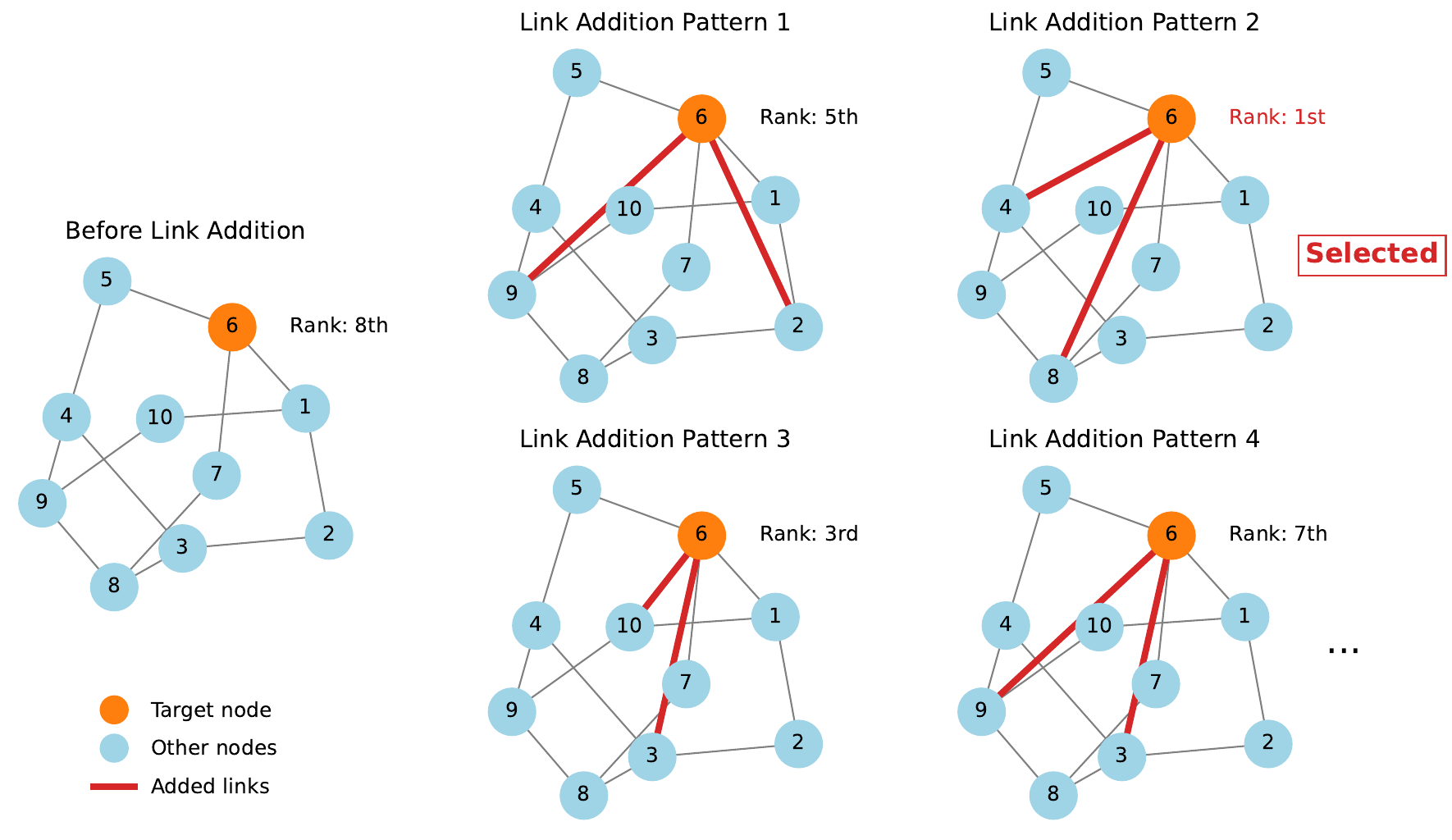}
\caption{Illustration of the Link Selection Problem for Influence Ranking Improvement (LSP-IRI). Different link addition patterns lead to different rank improvements for the same target node, and the goal is to select the link set that yields the largest improvement.}
\label{fig:irip}
\end{figure}

In this paper, to understand the feasibility of influence ranking modification, we tackle the Link Selection Problem for Influence Ranking Improvement (LSP-IRI).
Given a target user and a fixed link budget, the goal of LSP-IRI is to select a set of links that maximizes the improvement in the target user's ranking under an IM-based influence measure.
Figure~\ref{fig:irip} provides an illustrative overview of the LSP-IRI.
In many IM settings, nodes with more outgoing information-flow links tend to achieve larger influence spread.
Thus, increasing the number of outgoing information-flow links from the target user to other users, which corresponds to gaining additional followers, can be viewed as a practically motivated local intervention for increasing the target user's influence.
By studying such link additions, we aim not merely to improve a particular user’s ranking, but to assess how susceptible IM-based influence rankings are to limited, local structural modifications. 
Despite the simplicity of this objective, the underlying technical challenges are substantial.
First, the search space for link addition is inherently combinatorial: selecting a limited number of new connections from a large set of candidates leads to a combinatorial explosion, making exhaustive search infeasible in large-scale networks.
Second, evaluating the impact of each candidate modification is computationally expensive because exactly computing the expected influence spread under the IC model is \#P-hard~\cite{chen2010scalable}.
Furthermore, unlike traditional IM, which optimizes the absolute value of influence spread, our objective concerns a node’s relative ranking.
This requires not only estimating the target node’s influence but also accounting for how the rankings of all other nodes may change.
These challenges call for efficient heuristics and scalable estimation techniques that balance computational efficiency with ranking accuracy.

To address the LSP-IRI and systematically examine the feasibility of influence ranking modification, we examine two representative heuristic strategies with different computational characteristics.
The first method is a greedy method that iteratively selects links that maximize rank improvement.
This method is expected to achieve large ranking gains under small link budgets, but it incurs high computational cost because it repeatedly evaluates candidate links.
The second is a random search method that samples multiple candidate link sets and selects the best-performing one across several trials.
While this method yields comparatively moderate improvements, it significantly reduces computational overhead and remains scalable when the number of added links is large.
These methods allow us to analyze the trade-off between ranking improvement and computational efficiency, thereby providing a practical basis for assessing the susceptibility of IM-based influence rankings under different link budgets and computational constraints.

The main contributions of this study are summarized as follows:
\begin{itemize}
\item We formulate the LSP-IRI, which captures how the influence ranking of a target node can be changed through limited link additions under the IM framework. Beyond ranking optimization, this formulation provides a way to assess the robustness of IM-based influence rankings to small structural modifications.
\item We examine two complementary link selection strategies: a greedy method that directly optimizes rank improvement and a random search method that substantially reduces computational cost while maintaining scalability for larger link budgets.
\item We conduct experiments on multiple real-world networks and show that a small number of added links can substantially improve the ranking of target nodes. These results demonstrate the susceptibility of IM-based influence rankings to local structural changes and reveal the trade-off between ranking gain and computational efficiency.
\end{itemize}

\section{Related Work}

Influence maximization has been extensively studied as a framework for selecting influential seed nodes, with substantial efforts devoted to improving the scalability of influence estimation and seed selection.
The IM problem, introduced by Kempe et al.~\cite{kempe2003maximizing}, aims to find a seed set that maximizes expected influence spread under a diffusion model.
Because exact influence computation and repeated marginal-gain evaluation are computationally expensive, many studies have developed scalable IM algorithms, including greedy-based acceleration techniques and local-structure-based approximation methods~\cite{leskovec2007cost,goyal2011celfpp,chen2010scalable,chen2010ldag}.
Among scalable IM approaches, Reverse Influence Sampling (RIS) estimates influence spread using reverse-reachable (RR) sets rather than repeated forward diffusion simulations~\cite{borgs2014maximizing}.
An RR set is generated by sampling a node and tracing the diffusion process in reverse, and the influence spread of a seed set is estimated from the fraction of RR sets it covers.
In this study, we adopt RR-set-based estimation to compute IM-based influence rankings on the original graph and on graphs after link addition.

Recent studies have further extended IM-based techniques to practical influencer recommendation and optimization settings, but they still primarily address the platform-side task of selecting influential users.
For example, Dam et al.~\cite{dam2025fair2vec} studied fairness- and topic-aware influencer recommendation, and Biswas et al.~\cite{biswas2022moimp} investigated multi-objective influence maximization under varying-size solutions and constraints.
These studies show that IM-related techniques are increasingly being adapted to practical recommendation and decision-making settings.
However, they mainly focus on how a platform or system should select influential users under additional objectives or constraints.
In contrast, our work shifts the focus to the target-node perspective and examines how the relative ranking of a target node can be modified through limited link additions.

Link interventions in social networks have been studied for multiple purposes, including augmenting influence diffusion, suppressing misinformation, and manipulating centrality-based node rankings, but their effect on IM-based influence rankings remains insufficiently understood.
Chen et al.~\cite{chen2024link} studied link addition for increasing the overall influence spread from given seed nodes, whereas Bayiz and Topcu~\cite{bayiz2022countering} considered graph alterations for reducing misinformation diffusion.
Waniek et al.~\cite{waniek2021hiding} examined link addition strategies for manipulating rankings defined by centrality measures such as degree, closeness, and betweenness centrality.
Such centrality-based rankings primarily capture structural prominence rather than diffusion-based influence.
Taken together, these studies suggest that even limited structural interventions can induce measurable changes in diffusion outcomes and node rankings.
However, existing work has mainly focused on global diffusion maximization, misinformation suppression, or centrality-based ranking manipulation.
In contrast, our work investigates how limited link additions affect the relative ranking of a target node under an IM-based influence measure.

\section{Problem Formulation}

\subsection{Preliminaries}

We model a social network as a directed graph $G=(V,E)$, where $V$ is the set of nodes representing users and $E$ is the set of directed edges representing relationships such as following or information flow.
Each directed edge $(u,v)\in E$ indicates that information can propagate from node $u$ to node $v$, for example, when node $v$ follows node $u$ on a social media platform.

To model information diffusion, we adopt the Independent Cascade (IC) model~\cite{kempe2003maximizing}.
Under the IC model, each newly activated node is given a single chance to activate each of its inactive out-neighbors.
For each edge $(u,v)\in E$, the activation attempt from $u$ to $v$ succeeds independently with probability $p(u,v)$.
The diffusion process continues until no further activations are possible.

For a seed set $S \subseteq V$, let $\sigma_G(S)$ denote the expected number of activated nodes in $G$ under the IC model.
The IM problem introduced by Kempe et al.~\cite{kempe2003maximizing} is formulated as follows:

\begin{prob}[Influence Maximization (IM)]
Given a directed graph $G=(V,E)$, a diffusion model, and a seed budget $k$, the IM problem is to find a seed set $S \subseteq V$ with $|S|=k$ that maximizes the expected influence spread:
\begin{equation}
S_k^* \in \argmax_{S \subseteq V,\, |S|=k} \sigma_G(S).
\end{equation}
\end{prob}

\subsection{Link Selection Problem for Influence Ranking Improvement (LSP-IRI)}

We next formalize the problem of improving the relative influence ranking of a target node by adding a limited number of outgoing links from the target node.
Let $\mathcal{A}$ denote a given IM algorithm, and let $\mathcal{A}_k(G)$ denote the seed set of size $k$ returned by $\mathcal{A}$ on graph $G$.
To quantify the position of a node under $\mathcal{A}$, we introduce the influence ranking in Definition~\ref{def:influence-ranking}.

\begin{definition}[Influence Ranking]
\label{def:influence-ranking}
Given a graph $G$, a node $v \in V$, and an IM algorithm $\mathcal{A}$, the influence ranking of $v$ is defined as
\begin{equation}
r_{\mathcal{A}}(v,G)
=
\min \left\{ k \in \{1,\dots,|V|\} \mid v \in \mathcal{A}_k(G) \right\}.
\end{equation}
That is, $r_{\mathcal{A}}(v,G)$ is the smallest seed budget at which $v$ is selected by $\mathcal{A}$.
\end{definition}

Given a target node $v$, we define the set of candidate outgoing links from $v$ as
\begin{equation}
\mathcal{C}(v)
=
\left\{ (v,u) \in \{v\} \times V \mid u \neq v,\ (v,u)\notin E \right\}.
\end{equation}
In other words, $\mathcal{C}(v)$ consists of all non-existing outgoing information-flow edges from $v$, excluding self-loops.

For any feasible link set $L \subseteq \mathcal{C}(v)$, let
\begin{equation}
G_L=(V,E\cup L)
\end{equation}
denote the graph obtained by adding all links in $L$ to $G$.
To evaluate the effect of a feasible link set on the target node's ranking, we introduce rank improvement in Definition~\ref{def:rank-improvement}.

\begin{definition}[Rank Improvement]
\label{def:rank-improvement}
Given a graph $G$, a target node $v$, an IM algorithm $\mathcal{A}$, and a feasible link set $L \subseteq \mathcal{C}(v)$, the rank improvement of $v$ under $L$ is defined as
\begin{equation}
\Delta_{\mathcal{A}}(v,L;G)
=
r_{\mathcal{A}}(v,G)-r_{\mathcal{A}}(v,G_L).
\end{equation}
A larger value of $\Delta_{\mathcal{A}}(v,L;G)$ indicates a greater improvement in the relative influence ranking of the target node.
\end{definition}

Using the above definitions, LSP-IRI captures the task of selecting a limited number of new outgoing links from the target node so as to maximize its rank improvement under the IM framework.
It is formulated as follows:

\begin{prob}[Link Selection Problem for Influence Ranking Improvement (LSP-IRI)]
Given a directed graph $G=(V,E)$, a target node $v \in V$, a link budget $l$, and an IM algorithm $\mathcal{A}$, the LSP-IRI is to find a link set $L \subseteq \mathcal{C}(v)$ with $|L|=l$ that maximizes the rank improvement of $v$:
\begin{equation}
L^*
\in
\argmax_{L \subseteq \mathcal{C}(v),\, |L|=l}
\Delta_{\mathcal{A}}(v,L;G).
\end{equation}
\end{prob}

\section{Link Addition Methods for Influence Ranking Improvement}

Because exactly solving the LSP-IRI requires evaluating a combinatorial number of feasible link sets, we use RIS-based heuristic methods that approximate rank improvement with substantially fewer evaluations.
Recall that the objective is to find a link set $L \subseteq \mathcal{C}(v)$ with $|L|=l$ that maximizes the rank improvement $\Delta_{\mathcal{A}}(v,L;G)$ of a target node $v$, where the influence ranking is evaluated by a given IM algorithm $\mathcal{A}$.

In our algorithms, we estimate influence rankings using a RIS-based IM algorithm~\cite{borgs2014maximizing}.
More specifically, for each graph instance, we generate RR sets and use them to efficiently estimate influence spread and compute the seed sets returned by $\mathcal{A}$.
Since an exhaustive search over all feasible link sets is computationally infeasible, we adopt two heuristic strategies: a greedy method and a random search method.

\begin{mthd}[Greedy Method]
Starting from an empty link set $L_0=\emptyset$, the greedy method incrementally adds one link at a time. At step $t \in \{1,\dots,l\}$, let $L_{t-1}$ denote the set of links selected so far. The next link is chosen so as to maximize the rank improvement of the target node after the addition:
\begin{equation}
e_t^*
\in
\argmax_{e \in \mathcal{C}(v)\setminus L_{t-1}}
\Delta_{\mathcal{A}}(v,L_{t-1}\cup\{e\};G).
\end{equation}
The selected link is then added to the current solution, that is,
\begin{equation}
L_t \leftarrow L_{t-1}\cup\{e_t^*\}.
\end{equation}
This procedure is repeated until $|L_t|=l$, and the final output of the greedy method is denoted by
\begin{equation}
\hat{L}_{\mathrm{GR}} = L_l.
\end{equation}
\end{mthd}

The greedy method locally maximizes rank improvement at each step and is therefore expected to identify highly effective link additions, especially when the link budget is small. However, this advantage comes at a substantial computational cost. At each iteration, the method must evaluate all remaining candidate links in $\mathcal{C}(v)\setminus L_{t-1}$, and each evaluation requires recomputing the target node's influence ranking under the modified graph.
Let $M_v=|\mathcal{C}(v)|$ denote the number of candidate outgoing links from the target node.
Since the method evaluates up to $M_v$ candidate links at each of the $l$ steps, the total number of rank evaluations is $O(lM_v)$.
In the worst case, because $M_v \leq |V|-1$, this is $O(l|V|)$.

\begin{mthd}[Random Search Method]
The random search method generates multiple feasible link sets of size $l$ by random sampling. Specifically, at each trial $i \in \{1,\dots,n\}$, a candidate link set
\begin{equation}
L^{(i)} \subseteq \mathcal{C}(v), \qquad |L^{(i)}|=l,
\end{equation}
is sampled uniformly at random from all feasible link sets of size $l$, and its rank improvement $\Delta_{\mathcal{A}}(v,L^{(i)};G)$ is evaluated. 
After $n$ trials, the method selects the trial with the largest rank improvement:
\begin{equation}
i^*
\in
\argmax_{i\in\{1,\dots,n\}}
\Delta_{\mathcal{A}}(v,L^{(i)};G).
\end{equation}
The final output of the random search method is then given by
\begin{equation}
\hat{L}_{\mathrm{RS}} = L^{(i^*)}.
\end{equation}
\end{mthd}

Compared with the greedy method, the random search method substantially reduces the computational burden because it evaluates only $n$ sampled link sets, resulting in $O(n)$ rank evaluations in total. Although this method does not exploit marginal gains and may therefore miss high-quality solutions, it remains practical for larger link budgets and large-scale networks, where exhaustive candidate evaluation becomes prohibitively expensive.

Overall, the two methods provide complementary approaches for examining the trade-off between rank-improvement effectiveness and computational efficiency.

\section{Experimental Methodology}

\subsection{Datasets}

We use four real-world social networks selected from the datasets used by Chen et al.~\cite{chen2024link}.
GRQC, NetHEPT, and DBLP are co-authorship networks, while Epinions is a directed trust network derived from an online consumer review platform.
GRQC represents a co-authorship network in the field of general relativity and quantum cosmology, NetHEPT represents a co-authorship network in high-energy physics, and DBLP represents a large-scale co-authorship network in computer science.
For the undirected co-authorship networks, we extract the largest connected component (LCC), whereas for the directed Epinions network, we extract the largest weakly connected component (LWCC), to remove isolated nodes and focus on structurally meaningful regions of the networks.
Table~\ref{tab:network_stats} summarizes the statistics of the components used in our experiments.

\begin{table}
\centering
\caption{Statistics of the datasets after extracting the largest connected component for undirected networks and the largest weakly connected component for the directed network}
\label{tab:network_stats}
\begin{tabular}{lrrr}
\hline
Network & \#Nodes & \#Edges & Avg. Degree \\
\hline
GRQC  & 4,158   & 13,428    & 6.46 \\
NetHEPT  & 8,638   & 24,827    & 5.75 \\
Epinions & 75,877  & 508,836   & 13.41 \\
DBLP     & 317,080 & 1,049,866 & 6.62 \\
\hline
\end{tabular}
\end{table}

\subsection{Comparison Methods}

We compare three link addition methods: the greedy method, the random search method, and a naive random baseline. The greedy and random search methods correspond to Method~1 and Method~2, respectively.

The naive random baseline samples a feasible link set $L \subseteq \mathcal{C}(v)$ with $|L|=l$ uniformly at random. To mitigate the variance inherent in this stochastic baseline, we repeat the link sampling and ranking evaluation process 100 times and report the average performance.
The present evaluation compares two basic search strategies with an unoptimized random reference. The results should be interpreted as a proof-of-concept evaluation of LSP-IRI rather than as a comprehensive comparison with all possible link-selection heuristics.

For all methods, we use 10,000 RR sets to estimate influence spread and compute influence rankings.
This RR-set size is chosen as a practical computational setting, motivated by prior empirical observations that reducing the number of RR sets can substantially improve efficiency while maintaining reasonable solution quality~\cite{chen2024link}.
In the random search method, the number of sampled candidate link sets is set to $n=100$.

\subsection{Experimental Settings}

We adopt the Weighted Cascade (WC) model~\cite{kempe2003maximizing} to instantiate the propagation probabilities in the IC model.
For each directed edge $(u,v)$, where information propagates from $u$ to $v$, we set $p(u,v)=1/\deg^{-}(v)$, where $\deg^{-}(v)$ denotes the indegree of node $v$.
This setting reflects the assumption that a node with many incoming information-flow links is less likely to be activated by any single in-neighbor, and it has been used in prior IM studies~\cite{chen2024link}.
For undirected co-authorship networks, we treat each undirected edge as bidirectional and use the node degree in place of the indegree.

We vary the initial rank of the target node and the number of added links as experimental parameters.
For each dataset, the target nodes are selected according to their initial influence rankings.
Specifically, we consider nodes initially ranked 50 and 100, as well as nodes whose initial ranks correspond to the top 5\%, 10\%, 25\%, 50\%, and 75\% positions in the influence ranking.
The number of added links is set to $l \in \{3,10,30\}$.
By evaluating all combinations of these parameters, we analyze how the link budget and the initial rank of the target node affect influence ranking improvement.

\subsection{Evaluation Metrics}

We evaluate the effectiveness of each method using two metrics: rank improvement and rank improvement ratio.
The rank improvement $\Delta_{\mathcal{A}}(v,L;G)$ is defined in Section~3.
A larger value of $\Delta_{\mathcal{A}}(v,L;G)$ indicates a greater improvement in the relative influence ranking of the target node.
To compare results across different initial-rank settings and network sizes, we also use the rank improvement ratio (RIR), which normalizes rank improvement by the maximum possible improvement from the initial rank.
For a target node with initial rank $r_{\mathcal{A}}(v,G)>1$, RIR is defined as
\begin{equation}
\mathrm{RIR}_{\mathcal{A}}(v,L;G)
=
\frac{\Delta_{\mathcal{A}}(v,L;G)}
{r_{\mathcal{A}}(v,G)-1}.
\end{equation}
This metric represents the proportion of the maximum possible upward rank improvement achieved by link addition.
An RIR value close to one indicates that the target node moves close to the top of the ranking, whereas a value close to zero indicates limited improvement.
A negative RIR value indicates that the target node's ranking becomes worse after link addition.

In addition to effectiveness, we evaluate computational efficiency by measuring the single-execution runtime required for link addition and influence-ranking evaluation for each target node.
For the naive random baseline, we conduct 100 independent trials and report the average single-execution runtime over these trials.
This evaluation allows us to analyze the trade-off between ranking improvement and computational cost and to characterize the strengths of each method from both effectiveness and efficiency perspectives.

\section{Experimental Results}

\subsection{Rank Improvement}

\begin{figure}[!t]
  \centering

  \begin{subfigure}[t]{0.48\textwidth}
    \centering
    \includegraphics[width=0.93\linewidth]{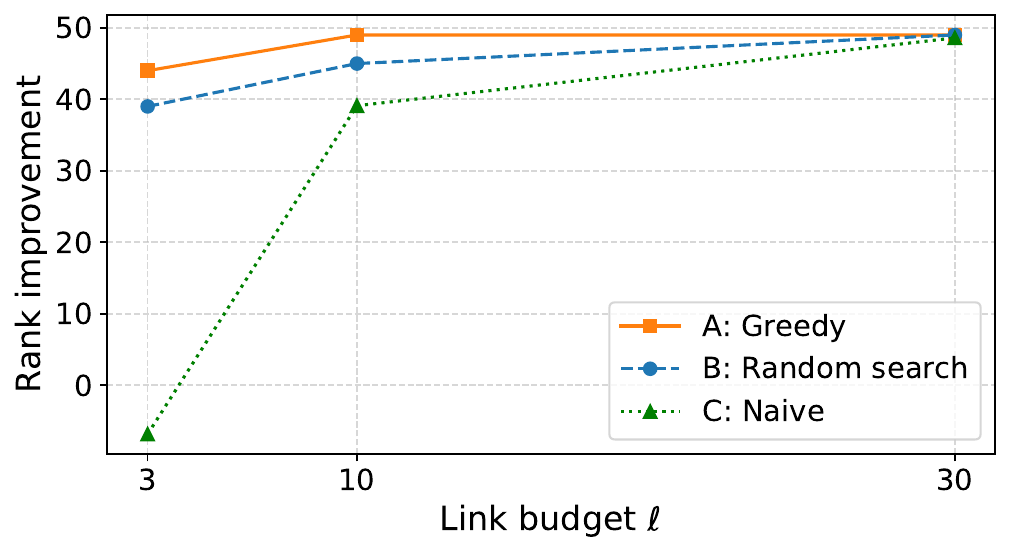}
    \caption{Initial rank: 50th}
  \end{subfigure}
  \hspace{0.02\textwidth}
  \begin{subfigure}[t]{0.48\textwidth}
    \centering
    \includegraphics[width=0.93\linewidth]{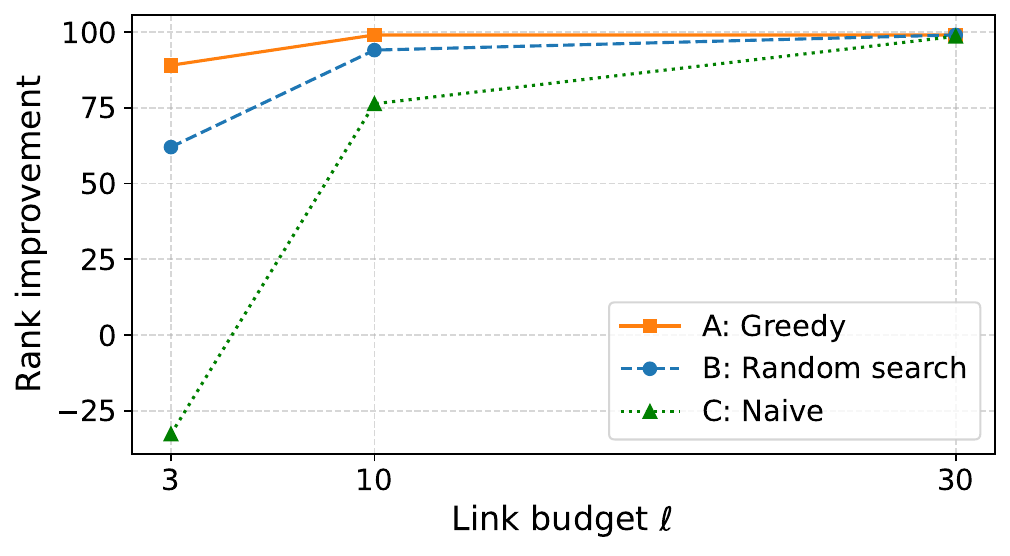}
    \caption{Initial rank: 100th}
  \end{subfigure}

  \vspace{0.5em}

  \begin{subfigure}[t]{0.48\textwidth}
    \centering
    \includegraphics[width=0.93\linewidth]{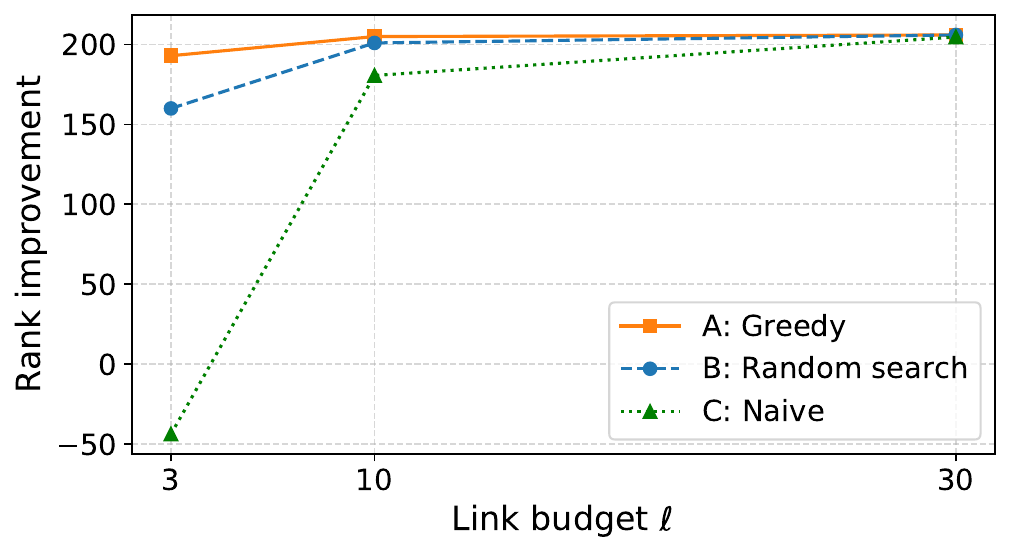}
    \caption{Initial rank: 207th (top 5\%)}
  \end{subfigure}
  \hspace{0.02\textwidth}
  \begin{subfigure}[t]{0.48\textwidth}
    \centering
    \includegraphics[width=0.93\linewidth]{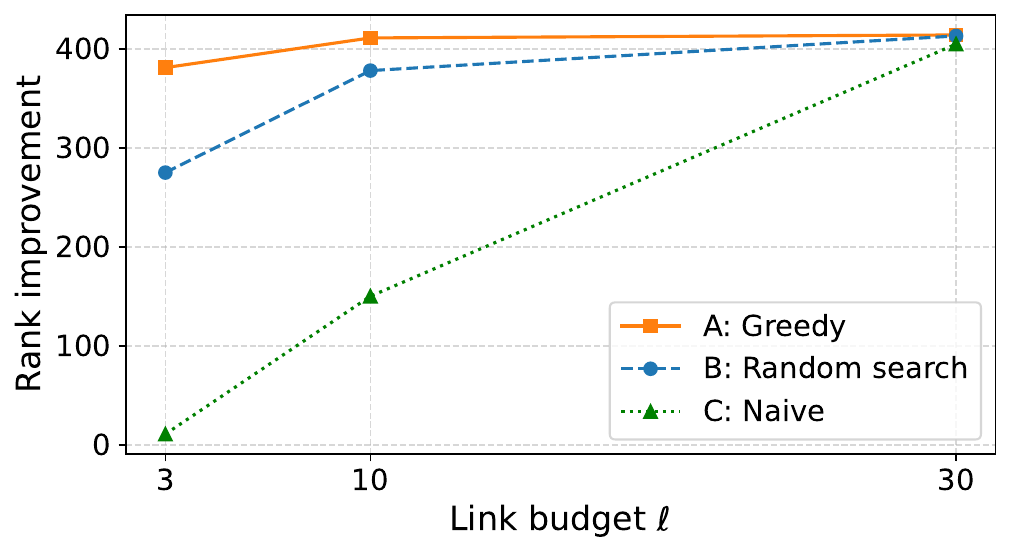}
    \caption{Initial rank: 415th (top 10\%)}
  \end{subfigure}

  \caption{Relation between link budget and rank improvement on GRQC. The greedy method achieves large improvements even with a small link budget, while the random search method tends to approach the greedy method as the link budget increases.}
  \label{fig:representative_rank_improvement}
\end{figure}

\begin{figure}[t]
  \centering
  \includegraphics[width=1\textwidth]{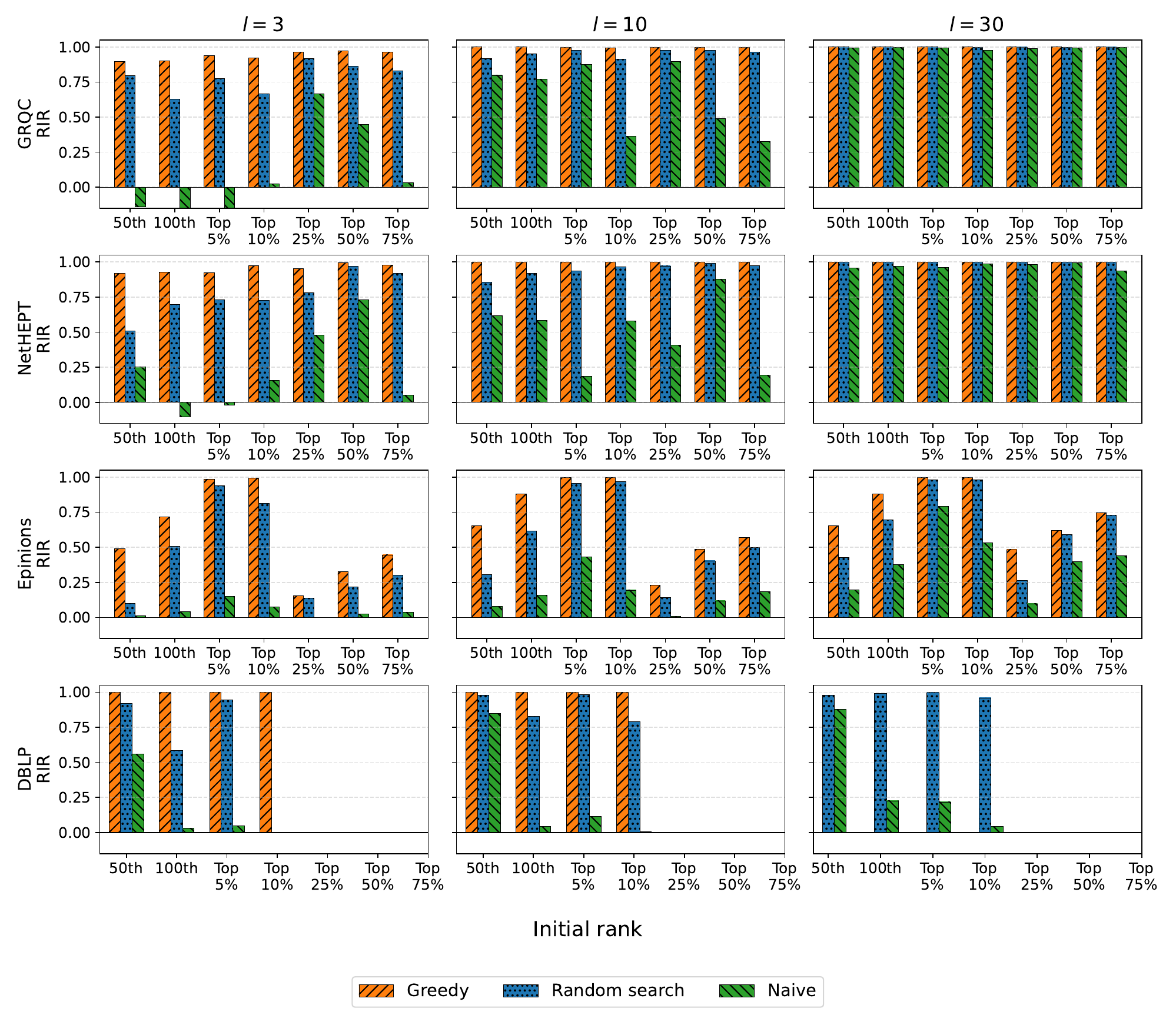}
  \caption{RIR across networks and link budgets. Each panel corresponds to a combination of a network and a link budget. The horizontal axis represents the initial-rank setting, and the vertical axis represents RIR. The greedy method generally achieves the highest RIR, while the random search method becomes competitive as the link budget increases.}
  \label{fig:rir_all_networks}
\end{figure}

We present the rank-improvement results using two complementary visualizations: raw rank-improvement plots for representative cases and RIR bar charts for the full set of evaluated experimental settings.
First, to illustrate the detailed behavior of the proposed methods, Figure~\ref{fig:representative_rank_improvement} shows the raw rank improvement $\Delta_{\mathcal{A}}(v,L;G)$ on GRQC for four upper-ranked target nodes, where the horizontal axis represents the number of added links and the vertical axis represents the resulting rank improvement.
Second, to provide a compact comparison across networks, link budgets, and initial-rank settings, Figure~\ref{fig:rir_all_networks} reports the RIR values defined in Section~5.
Each row corresponds to a dataset, each column corresponds to a link budget, and the bars compare the three link addition methods across different initial-rank settings.
For DBLP, some settings are omitted from the figure because they were not evaluated.
Specifically, the initial-rank settings corresponding to the top 25\%, 50\%, and 75\% nodes are omitted because many peripheral nodes have zero estimated influence, making fine-grained ranking beyond approximately the 40,000th position unreliable.
In addition, the greedy results for $l=30$ on DBLP are not reported because all evaluated target nodes reached rank 1 with $l=10$, making further greedy evaluations redundant.

Figure~\ref{fig:representative_rank_improvement} shows that the greedy method achieves the largest rank improvement in the representative GRQC cases.
The random search method provides moderate rank improvement, indicating that sampling-based link selection can still improve the target node's ranking.
In contrast, the naive random baseline shows unstable performance under small and moderate link budgets, indicating that unguided link addition does not reliably improve influence ranking.
As the number of added links increases, all methods tend to achieve larger improvements, and the performance gaps between the greedy method and the randomized methods become smaller in several settings.
Figure~\ref{fig:rir_all_networks} further shows that these tendencies are broadly consistent across different networks, link budgets, and initial-rank settings.
However, in larger networks such as Epinions and DBLP, lower-ranked target nodes do not always achieve substantial rank improvement.
This result suggests that mid- and high-ranked nodes often already have substantially more connections, and that adding even 30 links may be insufficient for lower-ranked nodes to close this connectivity gap.

\subsection{Computational Efficiency}

\begin{figure}[t]
  \centering
  \includegraphics[width=0.9\textwidth]{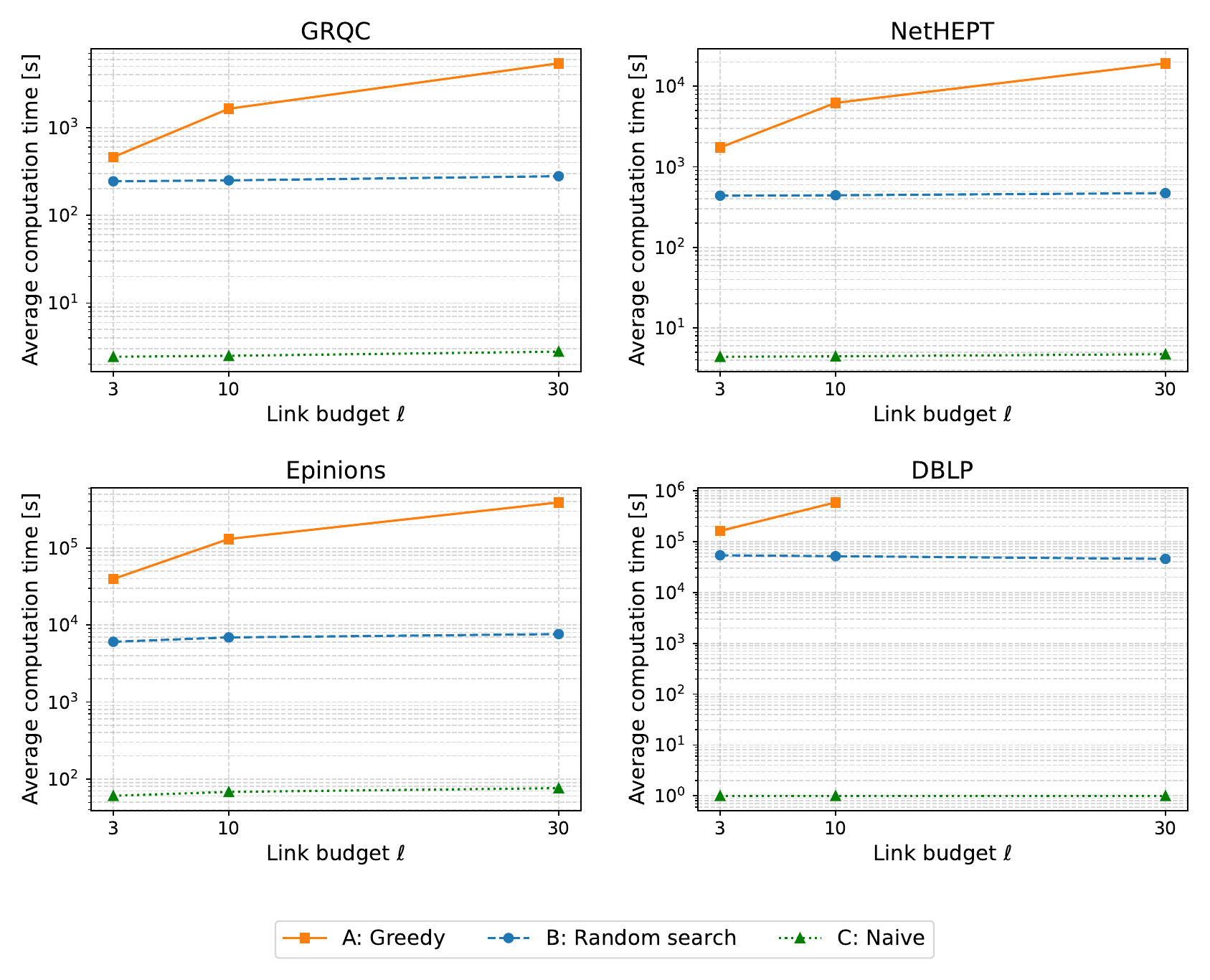}
  \caption{Runtime comparison across datasets and link budgets. Each panel corresponds to a dataset, and the vertical axis is plotted on a logarithmic scale. The greedy method incurs substantially higher runtime, especially for larger link budgets and larger networks, whereas the random search method maintains much lower computational cost by evaluating a fixed number of sampled link sets.}
  \label{fig:time_compare}
\end{figure}

We evaluate computational efficiency by measuring the average runtime required to select the link set for each target node.
Figure~\ref{fig:time_compare} compares the runtime of the three link addition methods across datasets and link budgets.
The vertical axis is shown on a logarithmic scale to make the large differences among methods visible.
Each panel corresponds to a dataset, and the horizontal axis represents the link budget $l$.

Figure~\ref{fig:time_compare} shows that the greedy method requires the longest runtime across all datasets.
Its runtime increases as the link budget grows, because it repeatedly evaluates candidate links at each step.
This tendency becomes more pronounced on larger networks, where the number of candidate links and the cost of ranking evaluation are higher.
In contrast, the random search method shows a much smaller increase in runtime because it evaluates only a fixed number of sampled link sets.
The naive random baseline achieves the lowest runtime, as it does not perform optimization during link selection.
These results show that the random search method provides a computationally efficient alternative to the greedy method while still achieving meaningful rank improvement in many settings.

\section{Discussion}

The experimental results provide two main implications for influence ranking improvement through link addition.

First, IM-based influence rankings can be substantially altered by a small number of local link additions.
This finding indicates that influence ranking is not merely a static outcome of the original network structure, but can be sensitive to limited local interventions.
From an individual-node perspective, this suggests that strategically forming a small number of connections can increase the likelihood of being selected as influential under an IM-based ranking mechanism.
From the perspective of algorithmic influence assessment, however, the same result reveals a potential vulnerability: influence rankings may be manipulated through relatively simple structural modifications.
Thus, the LSP-IRI formulation provides a useful framework not only for studying rank improvement, but also for assessing the robustness of influence-based ranking mechanisms.

Second, effective rank-improving links can be identified even under practical constraints such as limited link budgets and large network sizes.
In realistic social networks, a target user cannot freely acquire a large number of new connections, and exhaustive search over all candidate links is computationally infeasible.
Nevertheless, the greedy method achieves substantial ranking gains under small link budgets, indicating that carefully selected links can have a strong effect even when only a few additions are allowed.
At the same time, the random search method provides a computationally feasible alternative by evaluating only a limited number of sampled link sets while still achieving meaningful rank improvement in many settings.
Although the improvement may remain moderate when the link budget is small relative to the network scale or when the target node is initially ranked very low, these results suggest that influence ranking improvement can be realized under limited link budgets and computational constraints.

Several limitations should be considered when interpreting the empirical findings of this study, and they also suggest directions for future work.

First, the experiments were conducted on a limited number of real-world networks that are still smaller than very large online social platforms. Future work should evaluate the proposed framework on larger and more diverse networks to further examine its generalizability and scalability.

Second, this study does not provide theoretical guarantees for the proposed methods. Theoretical analysis is needed to clarify when substantial rank improvement can be expected and to characterize the behavior of the methods beyond empirical observations.

Third, the experiments are based on the IC model with the WC setting. Future studies should examine whether the findings remain robust under alternative diffusion models and propagation probabilities estimated from real behavioral data.

Fourth, our evaluation is limited to two heuristic strategies and a naive random baseline. Since existing link-intervention methods optimize different objectives, the results should be viewed as a proof of concept for LSP-IRI rather than a comprehensive benchmark.

Finally, the current formulation assumes that each candidate link has the same cost and can be added if selected. Incorporating heterogeneous link costs, link-formation feasibility, and acceptance probabilities would make the problem setting more applicable to practical scenarios.

\section{Conclusion}

This paper formulated the LSP-IRI, which captures how the influence ranking of a target node can be modified through limited link additions under the IM framework.
To address the LSP-IRI, we examined two complementary link selection strategies: a greedy method that directly optimizes rank improvement and a random search method that reduces computational cost.
Experiments on multiple real-world networks showed that a small number of added links can substantially improve the ranking of target nodes, demonstrating that IM-based influence rankings can be susceptible to local structural changes.
Overall, the LSP-IRI provides a framework for analyzing the extent to which IM-based influence rankings can be modified through limited local link additions.

\begin{credits}
\subsubsection{\ackname}
This work was partly supported by JSPS KAKENHI Grant No. JP25K03105 and JST ERATO Grant No. JPMJER2502.

\subsubsection{\discintname}
The authors have no competing interests to declare that are relevant to the content of this article.
\end{credits}
%
% ---- Bibliography ----
%
% BibTeX users should specify bibliography style 'splncs04'.
% References will then be sorted and formatted in the correct style.
%
% \bibliographystyle{splncs04}
% \bibliography{mybibliography}
%

\end{document}